\documentclass[11pt]{article}
\usepackage{amsfonts}
\usepackage{graphicx}
\usepackage{amsmath}
\makeatletter \@addtoreset{equation}{section}

\newcommand{\be}{\begin{equation}}
\newcommand{\ee}{\end{equation}}
\newcommand{\bea}{\begin{eqnarray}}
\newcommand{\eea}{\end{eqnarray}}
\begin{document}
\title{%
\begin{flushright}
 {\normalsize \small
GNPHE/0913 }
 \\[.5cm]
 \mbox{}
\end{flushright}
\textbf{ Supercordes,   Ph\'{e}nom\'{e}nologie et Th\'{e}orie-F } }
\author{ \hspace*{-20pt}
Adil Belhaj$^{1,3}$\thanks{\tt{belhaj@unizar.es}}, Leila
Medari$^{2}$
\\
[-6pt] \\
{\small $^{1}$  Centre National de l'Energie, des Sciences et des Techniques Nucl\'eaires,   Rabat, Morocco}\\
{\small $^{2}$ LPHEA, Physics Department, Faculty of Science
Semlalia,
Marrakesh, Morocco}\\
{\small $^{3}$ Groupement National de Physique des Hautes Energies, GNPHE, Si\`{e}ge focal: FSR}\\
[-6pt] {\small  Rabat, Morocco } } \maketitle
\begin{abstract}
Nous donnons une br\`eve id\'ee sur la  construction des mod\`eles
de jauge  en  th\'eorie de supercordes, tout particuli\`erement les
mod\`eles   avec quatre supercharges \`{a} quatre dimensions
provenant de la compactification de la
 th\'eorie-F. Nous discutons  la construction, selon Vafa, de la
th\'eorie-F  comme une  nouvelle approche de l'aspect  non
perturbatif du mod\`ele de la supercorde type IIB. Par la suite,
nous pr\'esentons des mod\`eles locaux de la th\'eorie-F, qui
peuvent g\'en\'erer de nouveaux mod\`eles de jauge  avec
supersym\'etrie $N=1$ \`{a} quatre dimensions pr\'esentant de grands
int\'er\^ets ph\'enom\'enologiques.

\end{abstract}
\thispagestyle{empty}
\newpage \setcounter{page}{1} \newpage

\section{Introduction}
 Dans les mod\`eles de supercordes,  une classe assez large  des
 th\'eories de jauge  est r\'ealis\'ee par des configurations de branes.
 On distingue deux approches:
 \begin{itemize}  \item {
  L'approche de Hanany-Witten \cite{HW}.}
  \item {
 L'ing\'enierie g\'eom\'etrique developp\'ee
  par Vafa et ses
 collaborateurs \cite{KKV}.}
  \end{itemize}
  Dans la premi\`ere  approche, les th\'eories des
 champs supersym\'etriques se pr\'esentent comme des
 th\'eories sur  le
 volume d'univers des branes qui englobe l'espace-temps. Alors que
 dans  l'ing\'enierie g\'eom\'etrique, les th\'eories des champs
 supersym\'etriques apparaissent comme une limite locale  de  certains
 compactifications. Dans cette  m\'ethode, les espaces de Calabi-Yau
 \`a  trois dimensions complexes sont r\'ealis\'ees localement comme  $K3\times
 B_2$, o\`u  $B_2$ est une base de dimension deux r\'eelles. Le  groupe de jauge  est
 obtenu imm\'ediatement de la singularit\'e de la fibre   K3 alors que la mati\`ere est
 g\'en\'er\'ee \`a  partir des choix non triviaux des  g\'eom\'etries de  la
 base $B_2$ .

 Actuellement, la m\'ethode  de l'ing\'enierie g\'eom\'etrique,   dans la compactification de la
  th\'eorie-F  vers  quatre dimensions,  est  consider\'ee  comme   l'un des moyens
   les plus prometteurs pour  obtenir des mod\`eles ph\'enom\'enologiques.
    Contrairement \`a  la plupart des mod\`eles  en th\'eories de supercordes,
       la compactification de la  th\'eorie-F peut   conduire \`a  une sym\'etrie de la th\'eorie
        de grande unification gr\^ace \`a  sa  structure riche en  branes et en flux.\\
        D'autant plus, l'un  des avantages de cette approche est sa relation
avec
   la th\'eorie de supercordes  het\'erotiques  compactifi\'ees  sur des espaces de Calabi-Yau admettant
           une  fibration elliptique. Cette  dualit\'e  sugg\'ere que certains
             mod\`eles  peuvent  \^etre  aussi  obtenus  \`a partir de la  comapactification
              de la th\'eorie-F poss\'edant encore plus d'avantages que
       la th\'eorie des supercordes h\'et\'erotiques.  Le  spectre des
                particules dans la th\'eorie-F est obtenu  \`a   partir de
                 la physique des  7-branes se propageant sur des g\'eom\'etries
   de  Calabi-Yau.

   R\'ecemment,  Beasley, Heckman et Vafa ont  construit une th\'eorie
    des champs  supersym\'etriques sur  des 7-branes intersectantes, pr\'eservant
      seulement  quatre supercharges \`a  quatre dimensions \cite{BHV}. Cette analyse  est bas\'ee
        sur  une g\'eom\'etrie locale  au voisinage  des 7-branes permettant
           d'\'etudier des  mod\`eles avec des sym\'etries de jauge  comme celles de
            la th\'eorie de grande unification. Ces mod\`eles de GUT
     (Grand Unified Theory en anglais)  sont une extension  \`a haute \'energie du mod\`ele standard
      de la physique des particules (les trois interactions \'elementaires \'electromagn\'etique,
       faible et forte, y sont d\'ecrites par un seul
 groupe de jauge  de type $SU_C(3)\times SU_L(2) \times U_Y(1)$).

 Le but principal  de ce travail est de donner une  br\`eve id\'ee sur la
 construction des mod\`eles de jauge  en th\'eorie des supercordes, tout particuli\`erement
 les mod\`eles $N=1$ \'e quatre dimensions provenant de la comapctification de la th\'eorie-F.\\
 Nous rappelons la
 construction, selon  Vafa,  de la th\'eorie-F comme une nouvelle approche
 de l'aspect non perturbatif du mod\`ele de  la supercorde type IIB\cite{Vafaf}.  Par la  suite,
 nous discutons des mod\`eles locaux de la th\'eorie-F,  qui peuvent
 g\'en\'erer  de nouveaux mod\`{e}les de jauge $N=1$ \`{a}
quatre dimensions d'inter\^{e}t ph\'{e}nom\'{e}nologique.

\section{Th\'{e}ories des Supercordes et la  ph\'{e}nom\'{e}nologie }
Rappelons que la th\'{e}orie des cordes est l'une des voies
envisag\'{e}es pour r\'{e}gler l'une des principales faiblesses du
Mod\`{e}le Standard (MS) de la physique des particules: l'unification des quatres interactions
\'{e}l\'{e}mentaires connues. En fait, dans ce cadre l\`{a}, on n'a pas pu disposer jusque l\`{a}, d'une th%
\'{e}orie quantique de la gravitation comme c'est le cas avec les
trois autres interactions fondamentales (\'{e}lectromagn\'{e}tique,
faible et forte).

 Lorsqu'on essaie de quantifier la gravitation,
des infinis apparaissent \`{a}
cause du caract\`{e}re ponctuel des particules\footnote{%
La physique des particules, mod\`{e}lise les particules
\'{e}lementaires comme des objets  ponctuels sans dimension.}, qu'on
ne peut \'{e}liminer par des processus standards de renormalisation.
Pour y remedier, en dessous d'une certaine \'{e}chelle ($<10^{-33}$
cm), les particules \'{e}lementaires sont consid\'{e}r\'{e}es comme des objets \'{e}%
tendus \`{a} une dimension spatiale poss\'{e}dant une tenison $T$,
appel\'{e}es cordes \cite{GSW,Vafal,P}. Plusieurs types de cordes
ont \'{e}t\'{e} consid\'{e}r\'{e}s: des cordes bosoniques (ouverte
ou ferm\'{e}e) ou encore des cordes fermioniques (ouverte,
ferm\'{e}e; chirale et non
chirale).

 A l'\'{e}chelle de Planck, les cordes bosoniques ferm\'{e}es sont assimil%
\'{e}es \`{a} des cercles; leurs excitations quantiques, contiennent
entre autre le graviton. Quant aux cordes bosoniques ouvertes, elles
sont assimil\'{e}es \`{a} des petits segments avec des conditions
aux bords de Dirichlet ou de Neumann; leurs excitations quantiques
contiennent naturellement les champs de jauge. La gravitation
quantique, mais aussi l'\'{e}lectromagn\'{e}tisme et les
interactions nucl\'{e}aires deviennent alors de simples
cons\'{e}quences de la g\'{e}om\'{e}trie et de la quantification du
mouvement de la corde.  Les deux types de cordes, ouvertes ou
ferm\'{e}es,  sont alors \`{a} la base de plusieurs mod\`{e}les
quantiques consistants. Ainsi le tout premier mod\`{e}le des cordes
\'{e}tait exclusivement bosonique, comprenant des cordes ouvertes et
ferm\'{e}es; impliquait un espace-temps \`{a} 26 dimensions. Mais
comme il n'impliquait pas les fermions, a vite perdu de son attrait
pour c\'{e}der rapidement la place aux mod\`{e}les dits de
supercordes, vivant dans un espace-temps \`{a} dix dimensions. On en
distingue 5 types:
\begin{itemize}  \item { Supercorde de type I
avec un groupe de  jauge $SO(32)$} \item { Supercorde
h\'et\'erotique   $SO(32)$}
\item { Supercorde
h\'et\'erotique   $E_8\times E_8$}
 \item{
Supercorde non chirale  type IIA} \item {Supercorde   chirale  type
IIB.} \end{itemize}
 Le  spectre des \'etats  non massifs de ces mod\`eles   contient,  le  dilaton
$\phi$,
  dont  $g_s =e^{\phi}$ est   la constante de  couplage de la th\'eorie, le
graviton $g_{ij}$  de spin 2, le  tenseur antisym{\'e}trique d'ordre
2 repr\'esent\'e par  le champ
    $ B_{ij} $ et    des  tenseurs  antisym\'etriques de  jauge $A_{\mu_1\ldots\mu_{p+1}}$
\begin{equation}
(g_{ij},\phi,B_{ij})\oplus A_{\mu_1\ldots\mu_{p+1}}.
 \end{equation}
Les tenseurs  $A_{\mu_1\ldots\mu_{p+1}}$ g\'en\'eralisant
     la  notion  de  potentiel  vecteur  $A_{\mu}$  \`{a}  des tenseurs
antisym\'etriques
      \'e  $p+1$ indices  (($p+1$)-formes,
$p=1,2,\ldots$),  se couplent  \`{a} des objets \'etendus  appel\'es
$p$-branes qui
       g\'en\'eralisent  la particule ponctuelle $(p=0)$ et la  corde  $(p=1)$.
          Les $p$-branes  sont
            des hypersufaces de $(p+1)$ dimensions,  dans l'espace-temps \'e
            dix dimensions, sur lesquelles  les extremit\'es des
cordes ouvertes  sont attach{\'e}es \cite{P}. Nous avons la
classification suivante des D-branes que l'on rencontre en th\'eorie
des supercordes
\begin{center}
\begin{tabular}{|c|c|c|c|c|c|}
 \hline
   & Type IIB & Type IIA & H\'et\'erotique  & H\'et\'erotique  & Type I \\
    &  &  & $E_8\times E_8$ & $SO(32)$ &  \\ \hline
  Type de corde & ferm\'ee & ferm\'ee & ferm\'ee & ferm\'ee & ouverte  \\
   & &  &  &  &  (et ferm\'ee) \\ \hline
  Supersym\'etrie& $N=2$ & $N=2$& $N=1$ & $N=1$ & $N=1$ \\
   de l'espace-temps & chirale  &  non chiral   &  &  &  \\ \hline
  Sym\'etrie de jauge &- & - &$E_8\times E_8$  & $SO(32)$ & $SO(32)$ \\ \hline
  D-branes & -1,1,3,5,7 & 0,2,4,6 &- & - & 1,5,9\\ \hline
\end{tabular}
\end{center}

 Pour ramener  ces th\'eories au
monde r\'eel, nous  avons besoin de les d\'efinir  dans notre espace
habituel \'e  1+3  dimensions. De ce fait, il faudrait compactifier
les six coordonn\'ees
 d'espace suppl\'ementaires.  Ainsi nous devons consid{\'e}rer des
g{\'e}om{\'e}tries  o\`u l'espace  de Minkowski
 {\'e} dix dimensions  $M^{1,9}$   se d{\'e}composant  en une vari{\'e}t{\'e} non
compacte correspondant  \'e l'espace-temps de Minkowski usuel
$M^{1,3}$ et une vari{\'e}t{\'e} compacte $ X^6$ de dimension $6$,
et de volume tr{\`e}s petit devant
 notre {\'e}chelle d'observation
 \begin{equation}
M^{1,9}\to M^{1,3}\times X^6.
 \end{equation}
La compactifications la plus \'etudi\'ee en  th\'eorie   des
supercordes est celle  qui concerne les  vari\'et\'es  de Calabi-Yau
de   dimensions trois complexes pr\'eservant le quart de
supercharges \'e  dix dimensions \cite{C}.  Ces derni\`eres restent
les candidates  les  plus probables   pour connecter les mod\`eles
de supercordes  au notre  monde  r\'eel. En g\'en\'eral, les
vari{\'e}t{\'e}s de  Calabi-Yau de dimension $3$ sont  des espaces
complexes, Kahl\'eriens, compacts  ayant un tenseur de Ricci nul et
un groupe d'holonomie $SU(3)$. Il existe diff\'erentes fa\c cons  de
construire  ces vari{\'e}t{\'e}s. Nous citons:

\begin{itemize}
  \item { Orbifolds de $T^{6}$}
  \item  {Fibration elliptique sur une vari{\'e}t{\'e} complexe   de
dimension $2$.}
  \item {Hypersurfaces dans les espaces projectives, o\`u plus
g{\'e}n{\'e}ralement dans les vari{\'e}t{\'e}s toriques.}
\end{itemize}
\`{A} basse \'{e}nergie, la compactification  de  la supercorde
h\'{e}t\'{e}rotique $E_{8}\times E_{8}$ sur une vari\'{e}t\'{e} de
Calabi-Yau peut donner une th\'{e}orie effective \'e quatre
dimensions avec la  supersym\'{e}trie $N=1$  de  type GUT. Ces
th\'{e}ories effectives ont un certain nombre de caract\'{e}ristiques int%
\'{e}ressantes; entre autres, nous citons
\begin{itemize}
  \item {Ils ont la structure des th\'{e}ories
supersym\'{e}triques de GUT}.
  \item {Chaque solution a un nombre d\'{e}fini de familles de
quarks et de leptons d\'{e}termin\'{e} par la topologie de l'espace
de Calabi-Yau.}
  \item {La sym\'{e}trie de jauge du mod\`{e}le standard peut
\^{e}tre int\'{e}gr\'{e}s dans un facteur $E_{8}$.}
\end{itemize}
Ces \'{e}volutions ont suscit\'{e} l'inter\^{e}t d'un grand nombre
des chercheurs  dans les ann\'{e}es $80$, mais  plusieurs questions
sont rest\'{e}es  cependant ouvertes, telles que:

\begin{itemize}
  \item {Choix de la vari\'{e}t\'{e} de Calabi-Yau}
\item { Les caract\'{e}ristiques  non perturbatives qui apparaissent  \`{a}
couplge fort}
\item { Le moyen de rendre massifs les champs scalaires (modules),
obtenus apr\`{e}s la compactification.} \end{itemize}

 Apr\`{e}s la deuxi\`{e}me r\'{e}volution de supercordes au milieu
des ann\'{e}es $1990$, ils sont apparu plusieurs nouvelles approches de ph\'{e}nom%
\'{e}nologie des particules en th\'{e}rie des supercordes.  Nous
citons par exemple:

\begin{itemize}
  \item { La
compactification des vari\'{e}t\'{e}s de Calabi-Yau avec des  flux.}
\item {
D-branes intersectantes  sur des g\'{e}om\'{e}tries singuli\`{e}res
(par exemple orbifolds et Orientifolds).}
 \item {La th\'{e}orie-M  vivant \`{a} $11d $. }

\item {La th\'{e}orie-F vivant \`{a} $12D $.} \end{itemize}
Nous rappelons dans de ce qui suivra quelques notions sur les deux
derni\`{e}res  th\'{e}ories.\\
\newpage
{\bf{Th\'{e}orie-M sur la vari\'{e}t\'{e} $G_{2}$}}\\
Selon Witten, la th\'{e}orie-M, qui est  la limite  de la
th\'{e}orie des supercordes type IIA  \`{a} couplage fort,  est
d\'{e}crite \`{a} basse \'{e}nergie par une th\'{e}orie de
supergravit\'{e} \`{a} $11$ dimensions\cite{W}.  Le supermultiplet
gravitationnel de ce mod\`{e}le
 comporte  le graviton, le gravitino de  Majorana et le tenseur
 3-forme. Le spectre de la th\'{e}orie-M contient \'{e}galement les
 solitons  M2-branes et M5-branes. Pour obtenir des
mod\`eles  supersym\'{e}triques  $N=1$,  \`{a} quatre dimensions, on
exige une compactification de la forme
 \begin{equation}
\mathcal{M}^{1,10}\to  M^{3,1} \times X^{7},
 \end{equation}
 o\`{u} $X^{7}$ est une vari\'{e}t\'{e} compacte de dimension sept
 dont le groupe d'holonomie  est $G_{2}$ \cite{J}. Rappelons que  le groupe $G_{2}$ est le plus petit
 des groupes de Lie complexes de type exceptionnel. Son alg\`ebre de Lie est de rang 2 et de dimension
 14. Sa repr\'{e}sentation fondamentale est de dimension 7. \\Il se trouve
si  la vari\'{e}t\'{e} de type  $G_{2}$ est reguli\`{e}re, la
th\'{e}orie effective resultante \`{a}  quatre dimensions n'a pas de
sym\'{e}trie de gauge non-abelienne. Pour surmonter ce probl\`{e}me,
il est n\'{e}cessaire
d'examiner la vari\'{e}t\'{e} de type  $G_{2}$ avec certains types de singularit\'{e}%
s \cite{AW}. Puisque cette vari\'{e}t\'{e} etant  de dimension
impaire,  elle  n'est  pas complexe. Le calcul  dans ce cas est
beaucoup plus difficile et  les techniques de l'analyse complexe ne
peuvent pas \^{e}tre utilis\'{e}es. Par cons\'{e}quent il n'y avait
pas beaucoup de progr\`{e}s dans la construction de  mod\`eles de
jauge provenant de la comapactification de la th\'{e}orie-M.

\section{La th\'{e}orie-F}
Rappelons que la th\'{e}orie-M d\'{e}crit la limite du couplage fort des th%
\'{e}ories de supercordes type  IIA et la supercorde
h\'{e}t\'{e}rotique $E_{8}\times E_{8}$. La question naturelle qui
se pose alors: quelle est la limite du couplage fort des supercordes
restantes?. Dans le cas de la th\'{e}orie type  IIB, il est
int\'{e}ressant  de rappeler tout d'abord qu'elle pr\'{e}sente une
sym\'{e}trie non perturbative $SL(2)$ agissant suivant ses
transformations modulaires sur la constante de couplage complexe
$\tau _{IIB}=\chi +ie^{-\phi }$ comme
\begin{equation}
 \tau _{IIB}\rightarrow \dfrac{a\tau _{IIB}+b}{c\tau
_{IIB}+d}; a,b,c,d\in \mathbf{Z}, ad-bc=1. \end{equation} Rappelons
que  le champ $\phi $ est le dilaton et le champ $\chi $ est l'axion
du secteur de \textbf{R-R.} Ce dernier se couple \`{a} la (-1)-brane, du spectre non perturbatif du mod%
\`{e}le  type  IIB, dont le dual magn\'{e}tique est une D7-brane
vivant \`{a} dix
dimensions. Dans la compatification du mod\`{e}le type IIB sur des espaces de Calabi-Yau, le dilaton $%
\phi $ et l'axion $\chi $ sont g\'{e}neralement consid\'{e}r\'{e}s
comme des constantes fixes. Cependant pour les compatifications
o\`{u} $\phi $ et $\chi $ ne sont pas
constantes, les configurations du vide peuvent naturellement \^{e}tre interpr%
\'{e}t\'{e}es dans le cadre d'une th\'{e}ories \`{a} 12 dimensions:
Th\'{e}orie-F \cite{Vafaf}. Dans ce mod\`{e}le, la sym\'{e}trie non
perturbative $SL(2)$ de la th\'{e}orie type  IIB est vue comme le
groupe des diff\'{e}omorphismes d'un tore $T^{2}$ formant avec
l'espace-temps du mod\`{e}le  de supercorde type IIB un espace \`{a}
12 dimensions. Selon Vafa,
 le champ complexe $\tau _{\substack{ IIB  \\
}}$ est donc identifi\'{e} avec la structure complexe d'un tore
suppl\'{e}mentaire $T^{2}$ \cite{Vafaf}. Dans ce cas, nous avons la
relation suivante
\begin{equation}
\tau _{IIB}=\chi +ie^{-\phi }=\tau \left( T^{2}\right).
\end{equation}
Cette identification montre que la th\'{e}orie de type IIB \`{a} dix
dimensions est \'{e}quivalente \`{a} la th\'{e}orie-F
compatifi\'{e}e sur le tore $T^{2}:$ \\
\qquad \qquad $Supercordes$ $IIB$ $\grave{a}$ $D=10$ \ \ $\equiv $ $\ Th%
\acute{e}orie-F$ $\ sur$ $T^{2}.$

Bien que la th\'{e}orie-F reste encore mal comprise, nous pouvons
construire des mod\`{e}les de supercordes dans des dimensions inf\'{e}%
rieures provenant de la th\'{e}orie-F \cite{MV}. L'id\'{e}e est
qu'au lieu de compactifier sur le tore $T^{2},$ nous compactifions
la th\'{e}orie-F sur une vari\'{e}t\'{e} $W_{n+1}$ de dimension
$n+1$ elliptiquement fibr\'{e}e. Localement $W_{n+1}$ est r\'{e}%
alis\'{e}e comme:
\begin{equation}
W_{n+1}\sim T^{2}\times B_{n},
\end{equation}
o\`{u} $B_{n}$ est une base complexe de dimension $n.$  Il s'en suit
alors que la th\'{e}orie-F sur $W_{n+1}$ peut \^{e}tre vue comme la
th\'{e}orie de type IIB sur $B_{n}$ dans laquelle le champ $\tau
_{IIB}$ varie sur $B_{n}.$ Nous allons voir plus tard que certaines
compatifications de la th\'{e}orie-F peuvent \^{e}tre aussi
\'{e}quivalentes \`{a} des compatifications de la th\'{e}orie des
supercordes h\'{e}t\'{e}rotiques \cite{Vafaf,Vafal,MV}

Nous pr\'{e}sentons ici, un exemple de la compactification sur  la
surface K3 elliptique. Dans cette construction, la surface K3 est
r\'{e}alis\'{e}e comme:
\begin{equation*}
T^{2}\left( R_{1},R_{2}\right) \times S^2
\end{equation*}
o\`{u} $R_{1}$ et $R_{2}$ sont deux rayons du tore $T^{2}$. La
r\'{e}alisation alg\'{e}brique dans $\mathbf{C}^{3}$ de la fibration
elliptique de K3  peut \^{e}tre d\'{e}duite \`{a} l'aide de celle du
tore $T^{2}$ d\'{e}finie par l'\'{e}quation de Weierstrass
\begin{equation}
y^2 =x^{3}+fx+g.
\end{equation}
Les coefficients complexes $\ f$ et $g$ param\`{e}trisent la
structure  complexe $\tau \left( T^2\right) $ du tore $T^2$. La
fibration elliptique de $K3$ est alors obtenue en prenant $\ f$ et
$g$
comme des fonctions analytiques en la coordonn\'{e}e locale $z$ de la base $=%
\mathbf{P}^{1}\sim S^{2},$ c.\`{a}.d :
\begin{equation}
y^2 =x^{3}+f\left( z\right) x+g\left( z\right)  \label{y2}
\end{equation}
$z$ $\in S^{2},$ $f\left( z\right) $ et $g\left( z\right) $
\'{e}tant deux polyn\^{o}mes en $z$ de degr\'{e} $8$ et $12$
respectivement. L'\'{e}quation (\ref{y2}) d\'{e}crit en chaque point
$z$ de $\mathbf{P}^{1}$
un tore $T%
{{}^2}%
$ dont la structure complexe $\tau \left( z\right) $ est
d\'{e}termin\'{e}e
par le rapport $\dfrac{f^{3}}{g%
{{}^2}%
}$ \`{a} travers la relation
\begin{equation}
J\left( \tau \left( T%
{{}^2}%
\right) \right) =\dfrac{4\left( 24f\right) ^{3}}{27g%
{{}^2}%
\left( z\right) +4f_{8}^{3}\left( z\right) }
\end{equation}
o\`{u} $27g%
{{}^2}%
\left( z\right) +4f_{8}^{3}\left( z\right) =\delta $ est dit le
discriminant de la fibration elliptique, qu'on exprime d'une
fa\c{c}on approximative  comme $ \tau \left( T^2 \right) \sim
\frac{1}{\delta }\sim i\frac{R_{1}}{R_{2}}$.  La fibration
elliptique de la surface  K3 a deux partcularit\'{e}s suivantes:\\
$\bullet $ La structure complexe de la fibre elliptique $T^2 $ varie
analytiquement comme fonction de $z;$ sachant $\tau =\tau \left(
z\right) $\\
$\bullet $ La fibre elliptique $T^2$ se d\'{e}g\'{e}n\`{e}re en $24$
points $z$, de la base $S^2$ associ\'{e}s avec $ \delta=0 $.\\
Dans ce cas,  la structure complexe $\tau \left( T%
{{}^2}%
\right) $ du $T%
{{}^2}%
$ tend vers l'infini. Ceci signifie qu'au voisinage des z\'{e}ros de $\Delta $ le cercle de rayon $%
R_{2}$ du tore $T%
{{}^2}%
$ se contracte en un point.\\ Partant de la d\'{e}finition de la
th\'{e}orie-F et en utilisant l'argument adiabatique, nous trouvons
que la th\'{e}orie-F sur la surface K3 elliptiquement \ fibr\'{e}
est \'{e}quivalente \`{a} la th\'{e}orie  type IIB compactifi\'{e}e
sur \textbf{P}$^{1}$. Ceci
implique que la structure complexe $\tau \left( z\right) $ est donn\'{e}%
e par:
\begin{equation}
\tau \left( z\right) _{\substack{  \\ }}=\chi \left( z\right)
+ie^{-\phi \left( z\right) }\sim \frac{1}{\delta \left( z\right) }
\end{equation}
Puisque $\delta $ a 24 z\'{e}ros $z_{i}^{0},i=1,...,24$, on a :
\begin{equation}
\delta \sim \prod \left( z-z_{i}\right).
\end{equation}
Cette \'{e}quation est interpr\'{e}t\'{e}e comme d\'{e}crivant 24
D7-branes de la th\'{e}orie  type  IIB transverses \`{a}
$\mathbf{P}^{1}$ et localis\'{e}es aux $z_{i}^{0}.$ Selon la
d\'{e}g\'{e}nerescence des $z_{i}^{0}$, la th\'{e}orie de jauge
$N=1$ \`{a} $D=8$ vivant sur ces $24$ configurations de D7-branes a
un groupe de jauge G contrainte comme $U\left( 1\right) ^{24}\subset
G\subset U\left( 24\right)$.

 Notons, on peut ainsi construire une
nouvelle classe des th\'{e}ories de jauge  de type ADE  \`{a} huit
dimensions par la compactification sur la surface K3, o\`{u} le
groupe de jauge est d\'{e}termin\'{e} par la classification des
singularit\'{e}s elliptiques de type  ADE. Cette \'{e}tude est aussi
valable pour les mod\`{e}les de jauge non simplement lac\'{e}es de
type  BCFG \cite{BSf}.

La  compactification de la th\'{e}orie-F sur  la surface K3
elliptique  est  duale \`{a} la th\'{e}orie de
supercorde h\'{e}t\'{e}rotique sur $T%
{{}^2}%
$ avec un fibr\'{e} principal $G$ $\subset \left( E_{8}\times
E_{8}\text{ ou }SO\left( 32\right) \right) \cite{Vafaf,Vafal}$.
Cette nouvelle dualit\'{e} est confirm\'{e}e par une \'{e}tude
approfondie de l'espace des param\`{e}tres de chaque th\'{e}orie.
Les modules
 de la m\'{e}trique de K3 elliptique param\'{e}trisent l'espace homog\`{e}ne \cite{Vafaf,Vafal,MV}
\begin{equation}
\dfrac{SO\left( 18,2,\mathbf{R}\right) }{SO\left( 18\right) \times
SO\left( 2\right) }\times \mathbf{R}^{+}\times \mathbf{R}^{+}.
\end{equation}
Ainsi cet espace des modules est identique \`{a} celui de la supercorde h%
\'{e}terotique compactifi\'{e}e sur un tore $T^2$\cite{N}. Les deux
facteurs $\mathbf{R}^{+}\times \mathbf{R}^{+}$ peuvent \^{e}tre
interpr\'{e}t\'{e}s comme la constante de couplage complexe de la
supercorde h\'{e}t\'{e}rotique. Par ailleurs le facteur
$\dfrac{SO\left( 18,2,\mathbf{R}\right) }{SO\left( 18\right) \times
SO\left( 2\right) }$ est l'espace des modules du r\'{e}seau de
Narain $\Gamma ^{(18,2)}.$ La sym\'{e}trie de jauge perturbative $\left( G\right) $ de la supercorde h%
\'{e}t\'{e}rotique $G$ $\subset \left( E_{8}\times E_{8}\text{ ou
}SO\left( 32\right) \right) $ est obtenue \`{a} partir de la
g\'{e}om\'{e}trie de  la surface K3 en exigeant que la fibration
elliptique  admet des
singularit\'{e}s de type G au point $z$ de la sph\`{e}re $S%
{{}^2}%
$ o\`{u} le discriminant $\delta \left( K3\right) $ est nul. Cette
\'{e}quivalence entre l'espace des modules de la th\'{e}orie-F
sur la surface K3 et celui des mod\`{e}les de supercordes $N=1$ sur $T%
{{}^2}%
$ est confirm\'{e}e par l'utilisation des dualit\'{e}s connues entre les mod%
\`{e}les de supercordes. En effet, la dualit\'{e} th\'{e}orie-F
supercordes h\'{e}t\'{e}rotique \`{a} $D=8$, si elle est vraie, doit
\^{e}tre pr\'{e}serv\'{e} aussi \`{a} six dimensions par
l'application de l'argument adiabatique, en cons\'{e}rvant la
totalit\'{e} des supercharges \`{a} six dimensions par une
compactification
suppl\'{e}mentaire sur un tore $T%
{{}^2}%
$. Partant de la supercorde h\'{e}t\'{e}rotique sur
$T^{4}=T^{2}\times T^{2}$
ou de la th\'{e}orie-F sur $K3\times T^{2},$ nous obtenons une th\'{e}%
orie supersym\'{e}trique $N=2$ \`{a} six dimensions. Cette th\'{e}orie est \'{e}quivalente \`{a}
la supercorde  de type IIA sur la vari%
\'{e}t\'{e} K3  mais vue comme un orbifold de $T^{4}$.

En utilsant l'argument adiabatique, nous pouvons compactifier
d'avantage la th\'{e}orie-F vers des dimensions inf\'{e}rieures
\`{a} huit dimensions \cite{MV}.  Ceci exige que les
vari\'{e}t\'{e}s de Calabi -Yau $W_{n+1}$ ont \`{a} la fois une
fibration elliptique sur $B_{n}$ et une fibration par la surface K3
sur une base $\widetilde{B}_{n-1}$ de dimension $n-1$.  Les valeurs
moyennes du vide des diff\'{e}rents champs de la th\'{e}orie IIB
compactifi\'{e}e sont d\'{e}termin\'{e}es par la structure complexe
de
la vari\'{e}t\'{e} de Calabi-Yau elliptique $W_{n+1}$. La r\'{e}%
alisation alg\'{e}brique de $W_{n+1}$ est obtenue \`{a} partir de
l'\'{e}quation (\ref{y2}):
\begin{equation}
y%
{{}^2}%
=x^{3}+f( z,\widetilde{B}_{n-1}) x+g( z,\widetilde{B}%
_{n-1}) \text{ \ \ \ \ \ \ \ }z\in \mathbf{P}^{1}
\end{equation}
Les fonctions $ f$ et $g$ \'{e}tant des fonctions analytiques en
$z,$ et en coordonn\'{e}es locales $z_{i}$, ($i=1,.....,n-1)$ de
$\widetilde{B}_{n-1}$. \\Cette compactification suppl\'{e}mentaire sur $%
\widetilde{B}_{n-1}$ agit aussi sur le volume d'univers de la
D7-brane. Pour des compactifications vers 6 dimensions, le volume
d'univers de la D7-brane doit contenir deux directions compactes.
Ceci veut dire que la D7-brane est enroul\'{e}e sur un 2-cycle
compact donnant lieu \`{a} une  D5-brane localis\'{e}e aux points
singuliers de la vari\'{e}t\'{e} $W_{3}$. Une compatification
suppl\'{e}mentaire sur un deuxi\`{e}me 2-cycle donne des D3-branes
d\'{e}crivant une th\'{e}orie de jauge \`{a} quatre  dimensions.

 Dans ce qui suit, nous illustrons la contsruction  g\'{e}om\'{e}trique des
 th\'{e}ories de jauge de supersym\'{e}trie  $N=1$ \`{a} quatre dimensions. Ces
 mod\`{e}les sont obtenus  par la compactification de la th\'{e}orie-F  sur
 des  vari\'{e}t\'{e}s de Calabi-Yau locales \`{a} quatre dimensions
 complexes. Dans cette r\'{e}alization, la base de la fibration par la surface
K3 est une
 surface complexe satisfaisant la condition
\begin{equation}
h^{1,0}=h^{2,0}=0
\end{equation}
qui est n\'{e}cessaire pour obtenir des mod\`{e}les $N=1$ \`{a}
quatre dimensions qui sont les plus int\'{e}ressants  pour des
raisons ph\'{e}nom\'{e}nologique \`{a} notre \'{e}chelle
d'observation.

 \section{Mod\`{e}les locaux de la th\'{e}orie-F}
Ces deux derni\`{e}res ann\'ees,  plusieurs groupes de recherche,
essaient de mettre au point des mod\`eles ph\'enom\'enologiques
issus de la F-th\'eorie. Nous citons, entre autres les travaux de
Donagi et Wijnholt \cite{DW} et Beasley, Heckman et  Vafa
\cite{BHV}. Dans cette nouvelle approche, on consid\`{e}re  des
mod\`{e}les dans lesquels nous pouvons d\'{e}finir une limite locale
de la th\'{e}orie-F. Cette restriction a pour but justement de
d\'{e}coupler la gravit\'{e}, ne laissant par cons\'{e}quent que les
interactions de jauge \`{a} quatre dimensions. Ceci se r\'ealiserait
si on impose que le volume des dimensions transversales des
$4$-cycles enroul\'{e}s des $7$-branes soit arbitrairement grand. Il
s'ensuit que
\begin{equation}
M_{GUT}/M_{Pl} \to 0.
\end{equation}
Les 4-cycles,   qui sont contractables, doivent \^{e}%
tre des vari\'{e}t\'{e}s de \textit{K\"{a}hler \`{a} }courbure
positive. Ils sont class\'{e}s enti\`{e}rement et sont donn\'{e}s
par des vari\'{e}t\'{e}s dites vari\'{e}t\'{e}s de \textit{del
Pezzo} (ou surfaces de \textit{del Pezzo} ), d\'{e}sign\'{e}s  par
$dP_{k}$. L'entier $k$ prend les valeurs $0\leq k\leq 8$. Le
diagrame de Hodge  de cette g\'{e}om\'{e}trie est donn\'{e} par
\def\m#1{\makebox[10pt]{$#1$}}
\begin{equation}
  {\arraycolsep=2pt
  \begin{array}{*{5}{c}}
    &&\m{h^{0,0}}&& \\ &\m{h^{1,0}}&&\m{h^{0,1}}& \\
    \m{h^{2,0}}&&\m{h^{1,1}}&&\m{h^{0,2}} \\
    &\m{h^{2,1}}&&\m{h^{1,2}}& \\ &&\m{h^{2,2}}&&
  \end{array}} \;=\;
  {\arraycolsep=2pt
  \begin{array}{*{5}{c}}
    &&\m1&& \\ &\m0&&\m0& \\ \m0&&\m{k+1}&&\m{0.} \\
    &\m0&&\m0& \\ &&\m1&&
  \end{array}}
\end{equation}

 Cet espace
est obtenu  par l'essoufflement de l'espace projectif  $CP^2$, o\`u
les $k$ points  sur $CP^2$ sont remplac\'es  par des sph\`{e}res
$S^2$. Il peut \^{e}tre obtenu \'{e}galement \`{a} partir de  la
surface Hirzebruch d'ordre zero $F_0$. Cette derni\`ere  peut
\^{e}tre vue comme une fibration triviale d'une sph\`{e}re $S^2$
fibr\'{e}e sur une autre sph\`{e}re $S^2$ ($F_0=S^2\times S^2$). Les
espaces de  del Pezzo
ont une relation \'{e}troite avec les alg\`{e}%
bres de Lie exceptionnelles $E_{k}$. L'id\'{e}e de base est que ces
espaces contiennent des 2-cycles dont les \'{e}l\'{e}ments de la
matrice d'intersection coincident avec ceux de la matrice  de Cartan
des alg\`{e}bres de Lie exceptionnelles $E_{k}$ \cite{V}.

Rappelons que nous nous limiterons, dans ce qui suivra \`{a}  
l'\'{e}tude de la th\'{e}orie-F  compactifi\'{e}e sur une classe des
g\'{e}om\'{e}tries de Calabi-Yau  \`{a} 4 dimensions complexes.
Cette compactification se  r\'{e}aliserait en  deux \'{e}tapes:  une
premi\`{e}re compactification  sur la surface elliptique  K3 suivi
d'une autre sur un espace de del Pezzo  vers  notre univers (1+3)
dimensions. Pour cette r\'{e}alisation, la g\'{e}om\'{e}trie, en
pr\'{e}sence des branes, conduit vers de nouveaux r\'{e}sultats
concernant les th\'{e}ories des champs supersym\'{e}triques. En
particulier,  la compactification de la th\'{e}orie-F peut
g\'{e}n\'{e}rer  des mod\`{e}les \`{a} quatre dimensions de
supersym\'{e}trie  $N=1$ avec
 des  groupes de sym\'{e}tries  de type   $SU(5)$ ou $SO(10)$  avec trois familles de particules.
 En fait dans le cadre de
 la th\'{e}orie-F, on admet que les sym\'{e}tries de jauge sont
 d\'{e}t\'{e}ctables, \`{a} partir de 8 dimensions,  les
champs de mati\'{e}re le sont \`{a} partir de six dimensions, alors
que  les interactions ont lieu \`{a} quatre dimensions\cite{BHV}.\\
 Dans cette \'{e}tude, chaque enroulement des 7-branes sur un 4-cycle d\'{e}termine
un groupe de jauge qui peut \^{e}tre  identifi\'{e} avec le groupe
de jauge de type \textit{GUT}. L'introduction de la mati\`{e}re est
obtenue \`{a}  partir de l'intersection des 4-cycles.  En
g\'{e}neral, l'intersection de deux 4-cycles est un 2-cycle qu'on
peut associer aux champs de la mati\`{e}re chirale, qui sont
donn\'{e}s par les modes z\'{e}ro de l'op\'{e}rateur de
\textit{Dirac}. Le nombre de ces modes z\'{e}ro d\'{e}termine le
nombre de familles des particules. La valeur du  couplage de
\textit{Yukawa} est donn\'{e} par  l'intersection de trois 2-cycles
en un m\^eme point.

 Dans l'\'{e}tude des mod\`{e}les supersym\'{e}triques de la  th\'{e}orie de grande unification (GUT)
       bas\'{e}e  sur la th\'{e}orie quantique des champs \`{a}
quatre dimensions, la brisure de la sym\'{e}trie  $SU(5)$ du GUT au
groupe de gauge du mod\`{e}le standard $SU(3)\times SU(2)\times
U(1)$ exige la mise en
place des champs de \textit{Higgs} appartenant \`{a} des  repr\'{e}%
sentations de $SU(5)$.  Sauf que ce choix de groupe a rencontr\'{e}
de nombreux probl\`{e}mes, limitant son efficacit\'{e}.
 De m\^{e}me
de la compactification de la supercorde h\'{e}t\'{e}rotique sur un
espace de Calabi-Yau \`{a} trois dimensions,  la m\'{e}thode
standard consiste \`{a}  associer des lignes de \textit{Wilson} non
triviales aux cycles noncontractibles, n'a pas conu grand
succ\`{e}s.

Justement l'un des points forts de l'\'{e}tude locale de la
th\'{e}orie-F est son approche de la brisure de la sym\'{e}trie.
Dans ce cadre, la sym\'{e}trie de jauge $SU(5)$ est obtenue \`{a}
 partir de la compactification de la th\'{e}orie-F sur  une vari\'{e}t\'{e} locale \`{a} quatre
 dimensions complexes. La forme g\'{e}om\'{e}trique de cette vari\'{e}t\'{e}  est donn\'{e}e par  la
 surface complexe locale de  type  ALE\footnote{ALE: Les espaces asymptotiquement localement euclidiens.} avec une singularit\'{e} $A_5$  fibr\'{e}e sur un espace
 de del Pezzo.

Dans cette compactification, il n'y a pas de cycles
noncontractibles; un autre m\'{e}canisme de brisure de la
sym\'{e}trie de jauge $SU(5)$ \`{a} un groupe de gauge du mod\`{e}le
standard est alors requis.

La proposition de \textit{Vafa} et ses collaborateurs est
d'introduire des flux non nuls. Pour ce faire, il faut choisir le
sous-groupe $U(1)$ de $SU(5)$ correspondant. Particuli\`{e}rement si on choisit $%
U(1)_Y$ \`{a} d'hypercharge faible, on obtient exactement la
structure de brisure de la sym\'{e}trie d\'{e}sir\'{e}e. Ce flux est
dit \textit{hyperflux} dans la nouvelle litt\'{e}rature de la
th\'{e}orie-F, tout en consid\'{e}rant qu'un $2-$cycle de la surface
de \textit{del Pezzo} pourrait porter ce flux \cite{S}.

Donc,le choix des flux d\'{e}pendrait des buts recherch\'{e}s. Il
est possible de s'arranger pour obtenir des multiplets de particules
dans les diff\'{e}rentes repr\'{e}sentations de $SU(5)$. Dans le cas
des quarks et des leptons on aura besoin des multiplets de la
repr\'{e}sentation fondamentale et antisym\'{e}trique de  $SU(5)$.

\section{ G\'{e}om\'{e}trie  hyper-K\"{a}hlerienne  et la th\'{e}orie-F}
Dans cette section,  nous \'{e}tudions  le r\^{o}le  de la
g\'{e}om\'{e}trie hyper-k\"{a}hlerienne  dans la compactification de
la th\'{e}orie-F sur des vari\'{e}t\'{e}s locales. En particulier,
nous approfondissons l'\'{e}tude des singularit\'{e}s
hyper-k\"{a}hleriennes dans  la th\'{e}orie des supercordes.  Outre
les solutions connues, nous obtenons  d'autres solutions permettant
de d\'{e}river des nouvelles th\'{e}ories supersym\'{e}triques $N=1$
\`{a} quatre dimensions par le biais d'une r\'{e}alisation
g\'{e}om\'{e}trique  de la th\'{e}orie-F bas\'{e}e sur les
alg\`{e}bres de Lie \cite{ABBS,ABBMS}. Ce proc\'{e}d\'{e} implique
des graphes de Dynkin.   Nous allons nous limiter \`{a} une classe
sp\'{e}ciale des vari\'{e}t\'{e}s de dimensions 8 r\'{e}elles
fibr\'{e}e par  la surface  elliptique K3. Il est utile  de fournir
des exemples. Nous consid\'{e}rons  alors un syst\`{e}me de jauge
$N=4$ \`{a} deux dimensions  avec un groupe de jauge $U(1)^r$ avec
$r+2$ hypermultiplets dont les vecteurs de charges sont $Q^a_i$
($a=1\ldots,r\; i=1\ldots,r+2$) \cite{ABBS}. Cette th\'{e}orie
clasique est parametris\'{e}e par les constantes de couplages et les
triplets de {\it Fayet-Iliopoulos} $\xi^a$. Pour un super-potentiel
nul, les D-terms s'ecrivent en fonctions des scalaires complexes
comme suit \be \label{sigma4} \sum_{i=1}^{r+2}Q_i^a[\phi_i^\alpha
{\bar{\phi}}_{i\beta}+\phi_i^\beta {\bar{\phi}}_{i\alpha}
]=\vec{\xi}_a \vec{\sigma}^\alpha_\beta, \ee o\`{u} $\vec\sigma$
sont les matrices de Pauli.   Pour chaque valeur de $a$, ces trois
\'{e}quations r\'{e}elles se
 transforment comme un triplet  de $SU(2)_R$  de l'alg\`{e}bre
 supersym\'{e}trique $N=4$  agissant sur les structures complexes. Apr\`{e}s
 avoir divis\'{e}  l'espace  des solutions  par les
 transformations
 de jauge du groupe $U(1)^r$, nous obtenons  une vari\'{e}t\'{e} de dimensions  8 r\'{e}elles que nous
 pouvons montrer qu'elle est hyper-K\"{a}hlerienne. Cette construction
 est connue sous le nom de quotient hyper-K\"{a}hlerien g\'{e}n\'{e}ralisant
  l'\'{e}tude des singularit\'{e}s K\"{a}hleriennes \cite{KKV}.  En utilisant ces
techniques, nous pouvons construire des   vari\'{e}t\'{e}s
r\'{e}elles de dimensions 8 qui peuvent
 \^{e}tre utiles pour la compactification de la th\'{e}orie-F.
 Dans le cas o\`{u}  les vecteurs de charges $Q^a_i$  co\"{\i}ncident  avec
 les matrices de Cartan des alg\`{e}bres de Lie, ces vari\'{e}t\'{e}s sont
 donn\'{e}es par des fibr\'{e}s cotangente sur des espaces de type  $F_0$. En particulier, elles
   sont r\'{e}alis\'{e}es   comme des
 surfaces K3 locales fibr\'{e}es sur une collection des espaces  $F_0$ group\'{e}s
 suivant les graphes de Dynkin des alg\`{e}bres de Lie.  Dans ce cas,
 nous avons trois  types de g\'{e}om\'{e}tries \cite{ABS}:
\begin{itemize}
  \item {
  G\'{e}om\'{e}trie associ\'{e}e aux
 diagrammes de Dynkin de type finie}
  \item {G\'{e}om\'{e}trie associ\'{e}e aux
 diagrammes de Dynkin de type affine}  \item { G\'{e}om\'{e}trie associ\'{e}e
 aux
 diagrammes de Dynkin de type ind\'{e}finie.}
\end{itemize}
Pour d\'{e}crire la  physique    des 7-branes enroul\'{e}es  sur ces
g\'{e}om\'{e}tries, nous utilisons  les r\'{e}sultats
 de l'ing\'{e}nierie g\'eom\'etrique utilis\'es dans la compactification de
la th\'eorie des supercordes. Prenons diff\'erentes configurations
des 7-branes  que nous   enroulons   sur une  collection des espaces
de type $F_0$, le groupe  de jauge total prend la forme suivante
\begin{equation} \label{G}
 G\ =\prod_{i} SU(N_i),
\end{equation}
o\`{u} $i$ est le nombre des noeuds du diagramme de Dynkin de
l'alg\`{e}bre de Lie  en question.\\Par cons\'equent, nous  avons
donc trois mod\`{e}les class\'es  par les alg\`{e}bres de Lie qui
sont d\'{e}finies par:
\begin{equation}
\label{A}
 \sum_{j}K_{ij}^{\left( q\right)}N_j\ =\ qM_i,\qquad  q=-1,0,1.
\end{equation}
Dans cette  \'{e}quation, $K^{(+)}_{IJ}$, $K^{(0)}_{IJ}$ et
$K^{(-)}_{IJ}$ sont  respectivement les  matrices de Cartan des
classes finie, affine et ind\'{e}finie \cite{ABS}. L'\'{e}quation
(\ref{A}) peut \^etre interpr\'{e}t\'{e}e comme une condition
d'annulation des anomalies dans  la th\'{e}orie-F \cite{ABBMS}.
Notons que  les $ N_j$ d\'{e}notent les couleurs et alors  que les
$M_i$ d\'{e}signent les contributions  de la mati\`{e}re
fondamentale. Ces nombres  ne peuvent pas \^etre arbitraires.

Il s'ensuit que nous avons deux mod\`{e}les \`{a} quatre dimensions.
Le premier mod\`{e}le implique la pr\'{e}sence de  la mati\`{e}re
bi-fondamentale   et l'absence  de  la mati\`{e}re fondamentale.  Le
second  en plus de la mati\`{e}re bi-fondamentale, et fondamentale,
poss\`{e}de une sym\'{e}trie suppl\'{e}mentaire de saveur
associ\'{e}e \`{a} cette derni\`{e}re.
\\Notons que le premier mod\`{e}le est associ\'{e} aux  diagrammes de Dynkin de type
affine.  L'\'{e}quation (\ref{A}) s'annule alors
\begin{equation}
 \sum_{j}K_{ij}^{(0)}N_j\ =0
\end{equation}
montrant l'absence de  la mati\`{e}re fondamentale.
 Le second  mod\`{e}le impliquant
l'introduction de la mati\`{e}re  fondamentale est
 obtenue \`{a} partir des mod\`{e}les bas\'{e}s sur   une
 collection des $F_0$ intersectantes  suivant les graphes de Dynkin
 des alg\`{e}bres de Lie
 ind\'{e}finies ou finies.\\
 Notons que, pour certains exemples de la g\'{e}om\'{e}trie
 hyperbolique\cite{ABBMS}, nous  pouvons
construire un mod\`{e}le avec un groupe de jauge $SU_C (3)\times
SU_L (3) \times SU (3)$  avec une sym\'{e}trie de  saveur
suppl\'{e}mentaire de type $SU (3)$.  L'id\'{e}al serait qu'on
puisse briser ce groupe vers le groupe de sym\'{e}trie du MS.
Rappelons n\'{e}anmoins que $SU_C (3)\times SU_L (3) \times SU (3)$
est un sous-groupe maximal de la sym\'{e}trie exceptionnelle $E_6$
apparaissant comme un groupe de jauge  dans les mod\`{e}les de GUT.

{\bf Remerciement}: \\Les auteurs de ce pr\'{e}sent travail tiennent
\`{a} exprimer leurs plus vifs remerciements  aux chercheurs: R.
Ahllaamara,   M. Asorey, L. Boya, J. L. Cortes, M. P. Garcia de
Moral, Y. Lozano, E. Saidi, A. Segui, avec qui nous avons eu des
\'{e}changes et des discussions tr\`{e}s enrichissantes.

\end{document}